\documentclass[fleqn,12pt,twoside]{article}
\usepackage{espcrc1}

\usepackage{graphicx}

\def\Vec#1{\mbox{\boldmath $#1$}}
\newcommand{\lsim}{\mbox{\raisebox{-0.6ex}{$\stackrel{<}{\sim}$}}\:}

\newcommand{\AmS}{{\protect\the\textfont2
  A\kern-.1667em\lower.5ex\hbox{M}\kern-.125emS}}

\title{
Energy loss of partons traversing a QGP fluid
}

\author{Tetsufumi Hirano\address{Physics Department, University of Tokyo,
Tokyo 113-0033, Japan}
and Yasushi Nara\address{RIKEN BNL Research Center,
Brookhaven National Laboratory, Upton, N.Y. 11973, U.S.A.
}
\thanks{Present address: Department of Physics, University of Arizona,
Tuson, Arizona 85721, U.S.A.}
}
       
\begin{document}

\maketitle

\begin{abstract}
We analyse the azimuthal correlation function for high $p_{\mathrm{T}}$
charged hadrons 
in Au+Au collisions at the RHIC energy.
By using a dynamical model in which hydrodynamics is combined with
explicit propagation of high $p_{\mathrm{T}}$ partons,
we study the effect of the intrinsic transverse momentum of
initial partons on the azimuthal back-to-back correlations.
\end{abstract}


\section{INTRODUCTION}

Measurements of high $p_{\mathrm{T}}$ hadrons
at the Relativistic Heavy Ion Collider (RHIC)
may provide insight into 
a quark gluon plasma (QGP) \cite{QMproc}.
Since jets have to traverse the excited matter, their spectra
should be changed in comparison with the elementary hadron-hadron data.
The energy loss of jets has been proposed as a possible
signal of the QGP phase in relativistic heavy ion collisions
~\cite{Gyulassy:1990ye}.
Over the past year, a lot of work have been devoted to study
the propagation of jets through
 QCD matter~\cite{BaierQM2002}.

Recently, hadronic transverse momentum distributions
in Au + Au collisions
at $\sqrt{s_{NN}}=130$ and $200$ GeV have been measured at RHIC and
found that the yield of hadrons is suppressed
 at high transverse momentum~\cite{phenix_pi0,star_highpt}.
These data are consistent with the picture in which
jets considerably lose their energy in hot/dense medium.
In addition, the magnitude of the away-side jets is found to be
significantly suppressed in comparison with the near-side jets
in central collisions
from the azimuthal correlation measurements
at high $p_{\mathrm{T}}$ \cite{star_btob}.
The disappearance of the away-side peaks in central collisions
implies that the one jet can escape from
medium and the other correlated jet is absorbed by medium \cite{Bjorken}.

\begin{figure}[tb]
\begin{center}
\includegraphics[width=70mm]{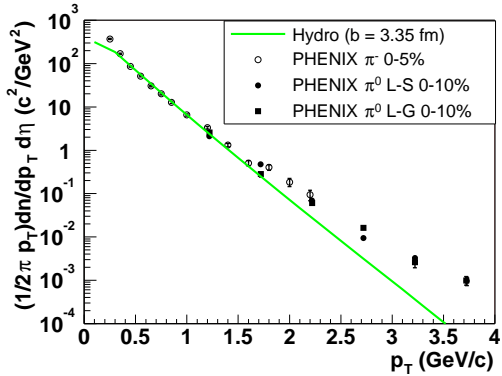}
\end{center}
\label{fig:DNDPT}
\caption{The transverse momentum spectrum from 
a hydrodynamic model including the effect of early
chemical freeze-out is compared with the compiled PHENIX data
in Au+Au collisions at $\sqrt{s_{NN}}=130$ GeV.}
\end{figure}

On the other hand, a novel hydrodynamical calculation, in which
the early chemical freeze-out is taken into account, show
that the $p_{\mathrm{T}}$ slope for pions is insensitive to
thermal freeze-out temperature $T^{\mathrm{th}}$
\cite{HiranoTsuda} and that the hydrodynamical result
starts to deviate from the experimental data
at $p_{\mathrm{T}} = 1.5$-$2.0$ GeV/$c$.
See Fig.~\ref{fig:DNDPT}.
This result naturally leads us to include
the mini-jets component in addition to the hydrodynamic component.

Thus, we developed the hydro+jet model in which a full three
dimensional space-time evolution of a fluid
is combined with
explicit propagation of jets \cite{HiranoNara}.
The hydro+jet model enables us to estimate the \textit{dynamical}
effect of hot matter on the parton energy loss in relativistic
heavy ion collisions.
By using this model, we study 
the azimuthal correlation functions at the RHIC energy.

\section{MODEL}
We have already had a hydrodynamic solution which reproduces
the experimental data in the low $p_{\mathrm{T}}$ region
($p_{\mathrm{T}} \lsim 1.5$ GeV/$c$)
at the RHIC energy.
For details on initialization of fluids,
used in this calculation,
see Ref.~\cite{HiranoTsuda}.

We include hard partons
(defined by particles that have transverse momentum greater
than 2 GeV/$c$ just after initial collisions)
 using pQCD parton model.
The spectrum of hard partons per hard collision
before traversing
hot matter can be written as
\begin{equation}
  E{d \sigma_{\mathrm{jet}}\over d^3 p}
  = K\sum_{a,b} \int g(k_{\mathrm{T,a}})d^2k_{\mathrm{T,a}}
                     g(k_{\mathrm{T,b}})d^2k_{\mathrm{T,b}}
  f_a(x_1,Q^2)dx_1f_b(x_2,Q^2)dx_2 E{d\sigma_{ab} \over d^3p},
\end{equation}
where $f$ and $g$ are the collinear and transverse parts of
parton distribution function, $x$ is the
fraction of longitudinal momentum
and $k_{\mathrm{T}}$ is the transverse momentum of initial partons
in a nucleon.
A factor $K$ is used for the higher order corrections.
The collinear parton distribution functions $f_a(x,Q^2)$
are taken to be CTEQ5 leading
order~\cite{cteq5}.
Gaussian distribution function $g$
for primordial transverse momentum $k_{\mathrm{T}}$
with the width of $\langle k_{\mathrm{T}}^2\rangle = 1$ GeV$^2/c^2$
 is assigned to the shower initiator
 in the QCD hard $2\to2$ processes.
We use PYTHIA 6.2~\cite{pythia} to simulate each hard scattering
and initial and final state radiations
in the actual calculation.
In order to convert hard partons into hadrons,
we use an independent fragmentation
model in PYTHIA after hydrodynamic simulations.
We have checked that this hadronization model with $K=2$
provides good agreement
with the transverse spectra of charged hadrons
in $p\bar{p}$ collisions~\cite{ua1}
and of neutral pions in $pp$ collisions above $p_{\mathrm{T}} = 1$ GeV/$c$
at $\sqrt{s}=200$ GeV \cite{Torii}.

The number of hard partons is assumed to scale with the number
of hard scattering which is estimated by using Woods-Saxon nuclear density.
The transverse coordinate of a parton is specified by 
a distribution being proportional to the number of binary collision.

We assume that partons move along a straight path
and lose their energy throughout
traversing partonic medium.
We here use the following formula for parton energy loss \cite{Levai}
\begin{equation}
\Delta E = C \int_{\tau_0}^{\tau_f}d\tau (\tau-\tau_0)
\rho(\tau,\Vec{x}(\tau))
\ln\left(1+\frac{2E}{\mu^2R_{\mathrm{Au}}}\right).
\end{equation}
Here $\rho$ is a thermalized parton density from hydrodynamic simulation.
The screening mass $\mu$ is taken to be a typical value 0.5 GeV.
This formula is relevant to the realistic heavy ion reactions
where the number of jet scatterings is small \cite{Levai} and
thermalized partons are dynamically evolved \cite{v2jet1}.
The coefficient
$C (=0.25)$ is chosen so that the suppression factor $R_{AA}$ at
$p_{\mathrm{T}} = 3$ GeV/$c$ becomes the same value as the one
for the incoherent parton energy loss model
discussed in the previous work \cite{HiranoNara}.

\section{RESULTS}
\begin{figure}[tb]
\begin{center}
\includegraphics[width=70mm]{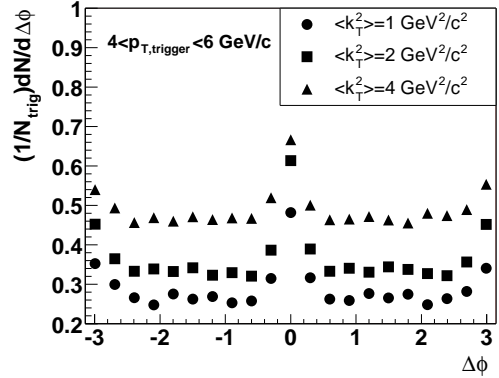}
\end{center}
\caption{Two-particle
azimuthal correlation function for charged particles
in Au+Au 200$A$ GeV central collisions.
Circle, square and triangle plots correspond to
$\langle k_{\mathrm{T}}^2 \rangle
= 1$, $2$ and $4$ GeV$^2$/$c^2$ respectively.}
\label{fig:BTOB}
\end{figure}

Figure \ref{fig:BTOB} shows the 
two-particle
azimuthal correlation function of
charged hadrons in midrapidity region ($\mid \eta \mid < 0.7$)
in Au+Au central collisions
at $\sqrt{s_{NN}}=200$ GeV. 
Here charged hadrons in $4 < p_{\mathrm{T, trigger}} < 6$ GeV/$c$ 
and in $2 < p_{\mathrm{T,associate}} < p_{\mathrm{T, trigger}}$
GeV/$c$
are defined to be trigger particles and associated particles
respectively.
We neglect the contribution from soft (hydro) components
which, in central collisions, does not affect the relative strength 
between the near-side peak ($\Delta \phi = 0$)
and the away-side peaks ($\Delta \phi = \pm \pi$). 
The away-side peaks do not vanish
even when the parton energy loss and the intrinsic transverse momentum
$\langle k_{\mathrm{T}}^2 \rangle=1$ GeV$^2$/$c$ are taken into account.
Since the initial multiple scattering inside colliding nuclei
releases the constraint of
exact back-to-back kinematics,
we estimate its effect on the azimuthal correlation function.
This effect can be parametrized phenomenologically 
by increasing $\langle k_{\mathrm{T}}^2\rangle$.
We also represent the results for $\langle k_{\mathrm{T}}^2 \rangle
= 2$ and $4$ GeV$^2$/$c^2$ in Fig.~\ref{fig:BTOB}.
The away-side peaks are found to be slightly weaken
as $\langle k_{\mathrm{T}}^2 \rangle$ increases.
Nevertheless, we can still see the away-side peaks.
Therefore our results indicates that,
in addition to parton energy loss in hot medium and
intrinsic $k_{\mathrm{T}}$ of partons in nuclei,
another mechanism is needed to obtain
the disappearance of back-to-back correlations

\section{SUMMARY AND DISCUSSION}
We have studied the effect of intrinsic $k_{\mathrm{T}}$ of partons
in a nucleus on the final azimuthal correlations of high $p_{\mathrm{T}}$
hadrons by using the hydro+jet model.
We found that 
the parton energy loss which is chosen so as to 
quantitatively reproduce the suppression factor $R_{AA}$ 
is insufficient for vanishing the back-to-back correlations
and that even a large $\langle k_{\mathrm{T}}^2 \rangle$
(= 4 GeV$^2$/$c^2$)
does not result in the disappearance of the away-side peaks.

The parton energy loss is related to the $p_\perp$ broadening
of jets in hot/dense medium \cite{Baier2}.
During propagation, jets can
get transverse momentum orthogonal to its direction of motion.
As a consequence, these partons follow zig-zag paths in hot medium.
This is naturally expected to affect the back-to-back correlation of jets.
The effect of $p_\perp$ broadening
on the azimuthal correlation function of high $p_{\mathrm{T}}$ hadrons
will be discussed elsewhere \cite{HiranoNara2}.

We would like to thank A.~Dumitru for motivating discussions
and continuous encouragement and I.~Vitev for valuable comments
on the formula of parton energy loss.

\end{document}